\documentclass[preprint,showpacs,showkeys,nofootinbib,superscriptaddress]{revtex4-1}
\usepackage{epsfig}
\usepackage{graphicx}
\usepackage{bm}
\usepackage{color}
\usepackage{empheq}
\usepackage{slashed}
\usepackage[all]{xy}

\newcommand{\beq}{\begin{equation}}
\newcommand{\eeq}{\end{equation}}
\newcommand{\bea}{\begin{eqnarray}}
\newcommand{\eea}{\end{eqnarray}}
\newcommand{\bean}{\begin{eqnarray*}}
\newcommand{\eean}{\end{eqnarray*}}
\newcommand{\Dslash}{D \kern -8pt \slash\,}

\begin{document}

\title{Ground state wave functions for the quantum Hall effect on a sphere and the Atiyah-Singer index theorem}

\author{Brian P. Dolan}

\email{bdolan@thphys.nuim.ie}

\affiliation{Department of Theoretical Physics, Maynooth University\\
  Maynooth, Co.~Kildare, Ireland}

\affiliation{School of Theoretical Physics\\
  Dublin Institute for Advanced Studies\\
  10 Burlington Rd., Dublin, Ireland}

\author{Aonghus Hunter-McCabe}

\affiliation{Department of Theoretical Physics, Maynooth University\\
  Maynooth, Co.~Kildare, Ireland}

\preprint{DIAS-STP-20-01}

\begin{abstract}

The quantum Hall effect is studied in a spherical geometry using the Dirac operator for non-interacting fermions in a background magnetic field, which is supplied by a Wu-Yang magnetic monopole at the centre of the sphere. Wave functions are cross-section of a non-trivial $U(1)$ bundle, the zero point energy then vanishes and no perturbations can lower the energy. The Atiyah-Singer index theorem constrains the degeneracy of the ground state.

The fractional quantum Hall effect is also studied in the composite Fermion model.
Vortices of the statistical gauge field are supplied by Dirac strings associated
with the monopole field. A unique ground state is attained only if the vortices
have an even number of flux units and act to counteract the background field, reducing the effective field seen by the composite fermions.  There is a unique gapped ground state and, for large particle numbers, fractions $\nu=\frac{1}{2 k+1}$ are recovered.

\end{abstract}

%\pacs{}
 
%\keywords{}

\maketitle

\section{Introduction}

The quantum Hall effects, both integer and fractional, have been a fascinating area of study ever since their first discovery.
Laughlin constructed trial ground state wave functions on the plane in \cite{Laughlin} and Haldane \cite{Haldane} considered a model of particles moving on the surface of a magnetic sphere ---
a sphere with a magnetic monopole at the centre.  It is common in such analyses to ignore the electron spin, since in strong magnetic fields it is assumed that electron spins are split and only the lower energy state is relevant to the problem so the spin can be ignored. In constructing the wave functions the particles are essentially considered to be spinless, but obey Fermionic statistics so that the many particle wave-function is anti-symmetric.
In particular the spin connection for spin-$\frac 1 2$ particles moving in a curved space plays no role in Haldane's construction.
Neither does the topology of the sphere play any real role, the sphere is merely a mathematical device that simplifies the analysis.

However the mathematics of Fermions on compact manifolds is very rich in both geometry and topology.  In particular the Atiyah-Singer index theorem
tells us that the Dirac operator on a sphere with a magnetic monopole at the centre has zero modes and an energy gap and constrains the number of positive and negative chirality zero modes.
This is true both relativistically and non-relativistically, the mathematics is essentially the same in both cases (the latter is basically the square of the former). 

There is a number of advantages in focusing on these zero modes: they are topological, for any fixed monopole charge they must always be there even if the magnetic field is distorted (provided the total magnetic charge does not change), and are therefore topologically stable, and they ensure that the zero point energy vanishes, there is an absolute minimum for the energy which no perturbation can reduce.
The number of zero modes is constrained by the index theorem, for a magnetic monopole of charge $m$ the difference between the number of positive chirality zero modes $n_+$ and the number of negative chirality zero modes $n_-$ is\footnote{There is a choice of sign on the right hand side which
depends on the definition of chirality, in our conventions it will be $-m$.}
\[n_+ - n_- =-\frac{1}{2\pi}\int_{S^2} F =-m\]
where $m$ is an integer (the first Chern class of a $U(1)$ bundle).
In particular if $m$ is positive $n_+$ can vanish in which case
\[n_-=m, \]
while if $m$ is negative $n_-$ can vanish and then
\[ n_+ = m.\]
When the zero modes have only one chirality
the number of linearly independent zero modes is exactly $|m|$ this reflects the degeneracy of the ground state.

In this paper we investigate the quantum Hall effect (QHE) on a sphere from the point of view of the Atiyah-singer index theorem and show how the zero modes relate to Haldane's version of the Laughlin ground state wave function. 
While the role of topology has long been appreciated in the quantum Hall effect to our knowledge the Atiyah-Singer index theorem has not been exploited to any great extent, except for the case of relativistic 4-component fermions in graphene, \cite{Park+Yi,Ezawa}, and in non-commutative geometry in the higher dimensional QHE \cite{Hasebe}.  The index theorem in our present context of ordinary 2-component non-relativistic electrons was used in \cite{Koma}, though there it was in the context of non-Abelian gauge fields on a torus and the zero modes were not constructed explicitly.
The fact that the filling factor is related to the Chern class of a $U(1)$ bundle over a torus, which is a Brillouin zone in $k$-space, was pointed out in \cite{NTW},
but while Chern classes are part of the index theorem for the Dirac operator, the theorem itself is much more than just Chern classes, in the context studied here it is about zero modes of the Dirac equation.  Another topological aspect of the quantum Hall effect is its relation to Chern-Simons theory but this is not relevant to the index theorem, Chern-Simons theories are only defined in odd dimensions and the index always vanishes in odd dimensions. 

The integer QHE is studied first, with a uniform magnetic flux through the surface of the sphere.  The exact ground state for $N$ non-interacting Fermions
is calculated and reproduces Haldane's result, equation (\ref{eq:Haldane}), for filling factor $\nu = 1$.

The fractional quantum Hall effect is then studied in the context of Jain's composite Fermion picture \cite{Jain}.
Magnetic vortices, represented by Dirac monopoles for which the Dirac string is viewed as a physical vortex of strength $v$ and is not a gauge artifact, are attached to the electrons.  The resulting composite particles move in the total magnetic field generated by the monopole plus the vortices.  For the wave function (a cross-section of a $U(1)$ bundle) 
to be free from singularities
the vortices necessarily have strength $|v|=2k$, where $k$ is
an integer, and act so as to reduce the strength of the uniform background field.  
Again zero modes can be constructed, equation (\ref{eq:fractional-ground-state}), 
and there is a unique ground state with an energy gap and for large $N$ the filling factor is $\nu = \frac{1}{2k +1}$.
This ground state can be related to Laughlin's ground state wave-function for the fractional QHE through a singular gauge transformation that removes the vortices.

The layout of the paper is as follows. In \S\ref{sec:Dirac} we review the Dirac operator on the surface of a sphere with
a magnetic monopole at the centre. In \S\ref{sec:ZeroModes} zero modes are constructed and shown to give a stable ground state
with an energy gap for filling factor $\nu=1$. For completeness wave-functions for energy eigenstates in the higher Landau levels are exhibited in terms of Jacobi polynomials in \S\ref{sec:HigherLL}.  Vortices are introduced and ground state wave functions for
the fractional quantum Hall effect are presented in \S\ref{sec:Vortices}. The results are summarized and conclusions presented in \S\ref{sec:Conclusions}.

\section{The Dirac operator on a sphere \label{sec:Dirac}}

\subsection{The Hamiltonian}

The full spectrum and eigenfunctions of the Dirac operator 
on a sphere in the absence of a magnetic monopole were studied in 
\cite{Abrikosov}.  On a magnetic sphere the spectrum can be derived from group theory \cite{Dolan}.  The eigenstates can be expressed simply in terms of Jacobi polynomials which were found to describe spinless particles on a magnetic sphere by Wu and Yang \cite{Wu+Yang}.

First consider a single non-relativistic spin-$\frac{1}{2}$ charged particle of mass $M$ confined to move on the surface of a sphere with a magnetic monopole at the centre of the sphere. The Hamiltonian is
\[ H = -\frac{\hbar^2}{2 M}{\Dslash^2}\]
where $i \Dslash$ is the (Hermitian) Dirac operator in the presence of the monopole and the sphere has unit radius.
This is bounded below and if there are zero modes of the Dirac operator they must be ground states with vanishing zero point energy.

The gauge potential for a monopole at the centre of the sphere is taken to be
\begin{equation*}A^{(\pm)}=\frac{m}{2} (\pm 1-\cos\theta) d \phi \qquad \Rightarrow \qquad
F=dA=\frac{m}{2}\sin\theta d\theta \wedge d\phi\end{equation*}
(the upper (lower) sign is for the upper (lower) hemisphere).
The monopole charge is
\[ \frac{1}{2\pi} \int_{S^2} F = m\]
with $m$ an integer.

We shall use a complex co-ordinate on $S^2$
\begin{equation*}
z=\tan\left(\frac{\theta}{2}\right) e^{i\phi},
\end{equation*}
in terms of which
\begin{equation}
  A^{(+)}(z) = \frac{i m}{2} \frac{(zd\bar{z} - \bar{z}dz)}{(1+ z \bar z)},
  \qquad
A^{(-)}(z) = \frac{i m}{2} \frac{1}{(1+ z \bar z)}
  \left( \frac{d z}{z} - \frac{d \bar z}{\bar z} \right)\label{eq:NS-gauge-transformation}
\end{equation}
and
\begin{equation}
  F = i m\frac{dz \wedge d\bar{z}}{(1 + z \bar z)^2}.
  \end{equation}

\subsection{The Dirac operator}

Choosing $\gamma^1 = \sigma^1$, $\gamma^2=\sigma^2$ the Dirac operator on the unit sphere is
\begin{equation} 
-i\Dslash = -i(1+z \bar z) \bigl(\sigma_+ D_z + \sigma_- D_{\bar z} \bigr)\label{eq:Dirac0},
\end{equation}
with
\begin{align}
D_z &=  \partial_z + \frac{(m + \sigma_3)}{ 2} \frac{\bar z }{ (1+ z \bar z)}, \label{eq:Dz}\\
D_{\bar z} &= \partial_{\bar z} - \frac{(m+\sigma_3)}{ 2} \frac{z }{ (1+ z \bar z)}\label{eq:Dbarz}
\end{align}
on the northern hemisphere (for electric charge $e=-1$).
More explicitly
\begin{equation}
\Dslash = \begin{pmatrix}0 & (1 + z \bar z)\partial_{z} + \frac{(m-1)}{ 2} \bar z  \\ 
(1 + z \bar z)\partial_{\bar z} - \frac{(m+1)}{ 2} z  & 0
\end{pmatrix},\label{eq:D-slash}
\end{equation}
which is anti-hermitian.

The curvature associated with the co-variant derivatives is
\[ [D_z,D_{\bar z}]= -\frac{(m+\sigma_3)}{(1+ z \bar z)^2}.\] 
The spin connection can be viewed as effectively increasing the magnetic charge by one for positive chirality spinors and decreasing it by one for negative chirality spinors.

\subsection{Angular Momentum}

The energy eigenstates can be classified by additional quantum numbers, in particular angular momentum will be a good quantum number but the definition involves some subtleties.
There are two aspects to the discussion of angular momentum: the presence of the
magnetic field and the orthonormal frame necessary to define spinors.
The former can be accommodated by defining
\[ L_a =  \epsilon_{a b}{}^c x^b ( p_c +  A_c)
= -i \epsilon_{a b}{}^c x^b ( \partial_c + i   A_c),\]
but in the presence of a magnetic field the algebra does not close, rather
\begin{equation}  
[L_a , L_b]=i \epsilon_{a b c}  \bigl( L_c  + e   x_c({\bf r}.{\bm B})\bigr).
\label{eq:JJB}
\end{equation}
In particular for a monopole field
\[ [L_a,L_b] = i \epsilon_{a b}{}^c \left(L_c - \frac{ m x_c}{2 r}\right),\]
but this can be countered by defining \cite{Fierz}
\[ J_a = L_a - \frac{ m x_a}{2 r},\]
giving a closed algebra
\begin{equation}
  [J_a,J_b] = i \epsilon_{a b}{}^c J_c.\label{eq:SU2-algebra}\end{equation} 
In terms of $z$,
\begin{align}
J_+ & = z^2  \partial_z + \partial_{\bar z} + \frac{m z}{1+ z \bar z},\nonumber\\
J_- & = -{\bar z}^2  \partial_{\bar z} - \partial_z + \frac{m \bar{z}}{1+ z \bar z},
\label{eq:Jm-def}\\
J_3 & = z \partial_z - \bar{z} \partial_{\bar z} + \frac{m}{2}.\nonumber
\end{align}
But this is not sufficient, Lie derivatives will also drag the orthonormal frame.
In the absence of any magnetic field the Lie derivative of a spinor $\psi$ with respect to a vector field $\vec L$ can be defined as \cite{SpinorLie}
\begin{equation}
 L^i D_i \psi + \frac 1 4 (d L)_{i j}\gamma^{i j}\psi \label{eq:Lie-psi}\end{equation}
where $\gamma^{i j} = \frac 1 2 (\gamma^i \gamma^j - \gamma^j \gamma^i)$ and
$d L$ is the exterior derivative of the 1-form metric dual to the vector $\vec L$ with $A_c$ set to $0$.
In terms of $z$,
\begin{align*} 
d L_+ &  = \frac{4 z}{(1 + z \bar z)^3} d z \wedge d \bar z\\ 
d L_- &  = - \frac{4 \bar z}{(1 + z \bar z)^3} d z \wedge d \bar z\\
d L_ 3 & =2 \frac{(1- z \bar z)}{(1+ z \bar z)^3} d z \wedge d \bar z 
\end{align*}
The prescriptions (\ref{eq:Jm-def}) and (\ref{eq:Lie-psi}) can be combined to give the Lie derivative of
a spinor in the presence of a magnetic monopole at the centre of the unit sphere 
in the following way
\begin{align}
{\bm J}_+ & = z^2 D_z + D_{\bar z} +\frac{(m+ \sigma_3) z}{1+ z \bar z} = z^2  \partial_z + \partial_{\bar z} + \frac{(m+ \sigma_3) z}{2},\nonumber\\
{\bm J}_- & = -{\bar z}^2  D_{\bar z} - D_z + \frac{(m+ \sigma_3)\bar{z}}{1+ z \bar z}=-{\bar z}^2  \partial_{\bar z} - \partial_z + \frac{(m+ \sigma_3)\bar{z}}{2},
\label{eq:J-def}\\
{\bm J}_3 & = z D_z - \bar{z} D_{\bar z} + \frac{(m + \sigma_3)}{2}\left(\frac{1- z \bar z}{1 + z \bar z}\right)
= z \partial_z - \bar{z} \partial_{\bar z} + \frac{(m + \sigma_3)}{2}\nonumber
\end{align}
on the northern hemisphere.
These satisfy
\begin{align} [{\bm J}_+,{\bm J}_-]&=2 {\bm J}_3, \qquad [{\bm J}_3,{\bm J}_\pm]= \pm {\bm J}_\pm\label{eq:SU2}\\
[{\bm J}_3,D_z] & =-D_z, \qquad [{\bm J}_3,D_{\bar z}]=D_{\bar z}, \nonumber\\
[{\bm J}_3,\Dslash] & =0,\qquad [{\bm J}^2,\Dslash]=0.\nonumber\end{align}
The square of the Dirac operator is related to the quadratic Casimir ${\bm J}^2$,
\[ -\Dslash^2 = {\bm J}^2 + \frac 1 4 (m^2 -1).\]
The eigenvalues of the Dirac operator on a coset space can be calculated from group theory, \cite{Dolan}. For the sphere
$S^2 \approx SU(2)/U(1)$, with a monopole at the centre, 
\[ \lambda^2 = n(n+|m|) \]
with degeneracy $2 n + |m|$,
where $n$ is a non-negative integer. Thus, with ${\bm J}^2 = J(J+1){\bm 1}$,
\[ J+\frac 1 2 = n + \frac {|m|}{2}.\]
There are zero-modes when $m\ne 0$ but when there is no background field
$n$ cannot vanish, in accordance with the Lichnerowicz theorem \cite{Lichnerowitz}.

\section{Zero modes \label{sec:ZeroModes}}

Spinors can be decomposed in terms of chiral eigenstates
\begin{equation*}\Psi = \begin{pmatrix}\chi_{+} \\ \chi_{-}\end{pmatrix}.\end{equation*}
Positive and negative chirality zero modes satisfy
\begin{align} D_{\bar z} \chi_+ &=0\\
D_{z} \chi_- &=0\end{align}
respectively.  From (\ref{eq:Dz}) and (\ref{eq:Dbarz}) it is immediate that,
on the northern hemisphere,
\begin{align}
\chi_+ & = {z^p}(1+ \bar z z)^{\frac{m+1}{2}}\label{eq:chi+p}\\
  \chi_- & = {{\bar z}^{\bar p}}(1+ \bar z z)^{-\frac{m-1}{2}} \label{eq:chi-barp}
\end{align}
satisfy these equations for any powers $p$ and ${\bar p}$.
However $p$ and $\bar p$ must be non-negative integers for $\chi_\pm$ to be well behaved at the north pole.
% $\chi_\pm$ are sections of a $U(1)$ bundle over the sphere and we do not want poles or branch cuts.  
We also want $\chi_\pm$ to be finite at the south pole, where
$|z|\rightarrow \infty$, so we must also require that 
 $\bar p - m +1 \le 0$ and $p + m +1 \le 0$.\footnote{At the upper limit of these bounds the magnitude of $\chi_\pm$ is finite but
the phase is undefined, this is a gauge artifact. A well defined phase is obtained at the south pole by performing the gauge transformation $\chi_\pm \rightarrow e^{i(m  \pm 1)\phi}\chi_\pm$ (the $\pm1$ arises from the spin connection).\label{footnote:phase}}
Thus, since $p$ and $\bar p$ are non-negative, negative chirality zero modes  require $0 \le \bar p \le m -1$ and 
positive chirality zero modes require $0\le p \le  -m-1$.
We see that for a negative chirality zero mode to exist it must be the case that $m\ge 1$ while a positive chirality zero mode
requires $m\le -1$.  There is no  value of $m$ for which there are both positive and negative zero modes. The index theorem then tells us that
\[ n_+ - n_- = -m \qquad \Rightarrow \qquad n_\pm = \pm m.\]
Thus for $m\ge 1$, $\bar p = 0,\ldots m-1$ exhausts the possibilities and for $m\le -1$, $p = 0,\ldots |m|-1$ exhausts the
possibilities.

The index theorem tells us that the number of zero modes here is $|m|$. This differs from Haldane's result that the degeneracy is $|m|+1$ and the difference is called the shift \cite{Wen+Zee}. In the multi-particle wave-function (discussed below) 
the shift is the difference between the number of flux quanta and the number of particles and it is non-zero in Haldane's analysis precisely because the electron spin and its coupling to the curvature of the sphere is ignored.  When electron spin and the spin connection on the sphere are treated properly the shift is zero and this is clearly shown here, it can be traced to 
the $(m+1)$ and $(m-1)$ terms in (\ref{eq:D-slash}), electrons with opposite spin couple to the spin connection with the opposite sign.  

The most general (un-normalised) zero modes are linear combinations of (\ref{eq:chi+p}) and (\ref{eq:chi-barp}) with constant co-efficients,
\begin{align}
\chi_+ & = \sum_{p =0}^{|m|-1} \frac{ a_{p} z^{p}}{(1 + \bar z z)^{\frac{|m|-1}{2}}}, \qquad \hbox{for}\ m\le -1,\label{eq:single-particle-chi+}\\
  \chi_- & = \sum_{\bar p =0}^{m-1} \frac{ a_{\bar p} {\bar z}^{\bar p}}{(1 + \bar z z)^{\frac{m-1}{2}}}, \qquad \hbox{for}\ m\ge 1.\label{eq:single-particle-chi-}
\end{align}

We shall analyze the $m < 0$ case (for positive $m$ simply complex conjugate the ground state wave functions).
The single particle ground state (\ref{eq:single-particle-chi+}) has degeneracy $|m|$,
which is a consequence of the index theorem.
%This is in contrast to the ground state for spinless particle constructed in \cite{Haldane} which, ignoring and overall normalisation and phase, has degeneracy 1.

The quantum Hall effect is a many particle phenomenon. Suppose we have $N$ identical particles on the sphere
and denote their co-ordinates by $z_i$, $i=1,\ldots,N$.
Ignoring interactions between the particles the Hamiltonian is
\[ H= -\frac{\hbar^2}{2 M} \sum_{i=1}^N \Dslash^2(z_i)\label{eq:H-N}\]
where
\[ \Dslash (z_i) = (1+z_i \bar z_i)
  \bigl(\sigma_+ D_{z_i} + \sigma_- D_{\bar z_i}\bigr).\]
The total ground state has zero energy and again consists of zero modes, but now for the zero mode associated with particle $i$ the co-efficients $a_p$ or $a_{\bar p}$ can be polynomials of the other $N-1$ co-ordinates.
The most general multi-particle ground state is
\begin{align*}
 \chi_+(z_1,\cdots,z_N) & = \left(\prod_{i=1}^N\frac{1}{(1 + \bar z_i z_i)^{\frac{|m|-1}{2}}}\right)\sum_{p_1, \ldots, p_N =0}^{|m|-1} 
                          a_{ p_1 \ldots p_N} {z_1}^{p_1} \cdots {z_N}^{p_N}.\\
\end{align*}
Since the particles are fermions the wave-function should be anti-symmetric, so
$a_{ p_1 \ldots p_N} $ should be anti-symmetric in its indices.  This requires $N\le |m|$ and leads to a degeneracy $\frac{|m|!}{N!(|m|-N!)}$. If $N>|m|$ then all the particles cannot fit into the ground state and some must go into the second landau level. If $N< |m|$ the
ground state is degenerate and cannot be expected to be stable under perturbations. There is a unique ground state, stable under small perturbations, if and only if $N=|m|$ in which case 
\begin{equation}
  \chi_+(z_1,\cdots,z_N)  = \left(\prod_{i=1}\frac{1}{(1 + \bar z_i z_i)^{\frac{|m|-1}{2}}}\right)\prod_{i<j} (z_i - z_j).\label{eq:Haldane}
\end{equation}
Thus there is a unique stable ground state if and only if the filling factor
\[\nu = \frac{N}{|m|}=1. \]
These are ground-state wave functions, the spherical versions of the Laughlin wave-functions on the plane for the integer quantum Hall effect.
For a sphere of radius $R$ the energy gap is
\[ \Delta E = \frac{(|m|+1)\hbar^2}{2 M R^2}=\frac{(N+1)\hbar^2}{2 M R^2}.\]
In the planar limit, $R\rightarrow \infty$, $N\rightarrow\infty$, keeping the
particle density $\rho = \frac{N}{4\pi R^2}$ finite, the energy gap
is 
\begin{equation}
  \Delta E = \frac{2 \pi  \rho \hbar^2}{M}= \frac{e B \hbar}{M},\label{eq:DeltaE}
\end{equation}
where $\frac{e B }{h} = \frac{|m|}{4 \pi R^2}$.

\section{Higher Landau levels \label{sec:HigherLL}}

When $N>|m|$ some particles must go into higher Landau levels. The energy
eigenfunctions in the higher Landau levels can be described by Jacobi polynomials,
$P^{(\alpha,\beta)}_n(\cos\theta)$. For spinless particles Jacobi polynomials were found in \cite{Wu+Yang}, for fermions Jacobi polynomials again appear
but the details differ due to the spin-connection \cite{Villalba}.

The eigenspinors of $-i\Dslash$ with  $\lambda_n = \pm \sqrt{n(n+|m|)}$, $n\ge 1$, are perhaps best exhibited using polar co-ordinates $(\theta, \phi)$ (on the northern hemisphere)
\beq
\psi_{\lambda_n,\alpha} = {\cal N}_{n,\alpha,\beta}
 \begin{pmatrix}
 \sqrt{n}\,z^\alpha \left(\cos \frac {\theta}{2} \right)^{|m|-1} P_n^{(\alpha,\beta)}(\cos\theta)\\ 
\mp i \bigl(\sqrt{n+|m|}\,\bigr) z^{\alpha+1} \left(\cos \frac{\theta}{2}\right)^{|m|+1}  P_{n-1}^{(\alpha+1,\beta+1)}(\cos\theta)
\end{pmatrix}\label{eq:sphere-eigenspinors}
\eeq
where $\alpha =-n,\ldots n+|m|$ labels the $2n+|m|$ independent degenerate states and $\beta$ is fixed by $\alpha+\beta = |m|-1$ and
\[{\cal N}_{n,\alpha,\beta}^2 =\frac{(2 n + |m|)}{8 \pi}\frac{\Gamma(n+1)\Gamma(n+|m|+1)}{\Gamma(n+\alpha+1)\Gamma(n+\beta+1)}\]
  a normalisation constant.\footnote{The eigenspinors are associated with irreducible representations of $SU(2)$ and are also expressible as Wigner $d$-functions.
  The full degenerate set for a given $n$ constitute a single column of the matrix in the $2 n+|m|$ dimensional irreducible representation of SU(2).}

The higher Landau levels now have both chiralities at the same energy level, but these can be separated by adding a Zeeman splitting
term $\mu |m| ({\bf 1} - \sigma_3)$ to the Hamiltonian.

The second Landau level corresponds to $n=1$ and has degeneracy $|m|+2$.
By the same argument as before the anti-symmetrized ground state multi-particle wave function is degenerate unless
\[ N= |m| + (|m|+2) =2(|m|+1) ,\] in which case the filling factor is
\[\nu = \frac{N}{|m|} = \frac{2 N}{N-2}\
  \rightarrow \ 2 \quad \mbox{as} \quad N \rightarrow \infty. \]
The resulting wave-function is non-degenerate, it is stable under perturbations and the energy gap between the  second and third Landau levels is
\[ \Delta E = \frac{\hbar^2}{2 M R^2}\bigl( 2(|m|+2) - (|m| +1)\bigr)
  = \frac{(|m|+3)\hbar^2}{2 M R^2}  = \frac{(N+4)\hbar^2}{4 M R^2}.\]

Repeating the argument for larger, but finite, $n$ we recover the integer quantum Hall effect in the limit of large $N$. There is a unique stable ground state
when the $n$-th Landau level is fully filled
\[ N=\sum_{k=0}^n (2 k + |m|)= (n+1)(n+|m|)\]
so
\[ \nu = \frac{N(n+1)}{N-n(n+1)} \ \rightarrow\ n+1  \quad \mbox{as} \quad N \rightarrow \infty. \]
The energy gap between level $n$ and level $n+1$
is
\[\Delta E  = \frac{\hbar^2}{2 M R^2}\bigl( (n+1)(|m|+n+1) - n(|m| +n)\bigr)
  =  \frac{(|m| + 2 n +1)\hbar^2}{2 M R^2}
=  \frac{\bigl(N -n(n+1)\bigr)\hbar^2}{2 (n+1)M R^2}.\]
In the planar limit
\[ \Delta E \ \rightarrow\  \frac{|m| \hbar^2 }{2 M R^2} = \frac{ e B \hbar}{M},\]
as expected.

\section{Fractional filling fractions \label{sec:Vortices}}

Fractional filling fractions in the quantum Hall effect can be understood in terms of flux attachment \cite{Jain}. A statistical gauge field is introduced
and the effective degrees of freedom are composite objects consisting of
electrons bound to statistical gauge field vortices.
These vortices then reduce the effective field seen by the composite fermions.

The gauge potential describing a uniform flux through the sphere arising from a monopole with charge $m$ at the centre of the sphere together
with $N-1$ vortices each of strength $v$ piercing the sphere at points $z_j$ is described in the appendix, (\ref{eq:A+}) and (\ref{eq:A-}). The gauge potential is 
\begin{align} 
A^{(+)}  & = \frac{v}{2 i} \sum_{j=1}^{N-1}\left( \frac{d z }{z-z_j}
-  \frac{d \bar z}{\bar z-\bar z_j}\right)
+ \frac{i m}{2}\left(\frac{z d \bar z - \bar z d z}{1+z \bar z}\right),\label{eq:A+a}\\
A^{(-)} 
&= \frac{v}{2 i} \sum_{j=1}^{N-1}\left( \frac{d z }{z-z_j}
-  \frac{d \bar z}{\bar z-\bar z_j}\right)
+ \frac{i}{2} \left( \frac{m}{(1+ z \bar z)}  +  (N-1) v \right) 
  \left( \frac{d z}{z} - \frac{d \bar z}{\bar z} \right).
\end{align}
The field strength is
\[ F = i\left(2 \pi v \sum_{j-1}^{N-1} \delta(z-z_j) + 
\frac{m}{2 (1+z \bar z)^2}\right) dz \wedge d \bar z.\]

The spectrum of the Dirac operator can be determined when there are magnetic vortices threading through the surface of the sphere in addition to a monopole at the centre. 
Omitting the self-energy of a composite fermion with its own vortex, and assuming all the vortices have the same strength $v_i=v$, the Dirac operator (\ref{eq:Dirac0}) on the northern hemisphere then involves covariant derivatives
\begin{align}
D_z & = \partial_z
+ \frac v 2 \sum_{j=1}^{N-1}\frac{1}{z - z_j}
+ \frac{( m +\sigma_3)}{2} \frac{\bar z}{1+z \bar z }
,\label{eq:Dzv}\\
    D_{\bar z} 
& = \partial_{\bar z}
- \frac v 2 \sum_{j=1}^{N-1}\frac{1}{\bar z - \bar z_j}
- \frac{( m +\sigma_3)}{2} \frac{ z}{1+z \bar z }.\label{eq:Dbarzv}
\end{align}
Using the identities 
\[ \partial_z\left(\frac{1}{\bar z -\bar z_i}\right)= \partial_{\bar z}\left(\frac{1}{z - z_i}\right)= 2\pi \delta(z - z_i)\] 
the commutator is
\[ [D_z,D_{\bar z}] = -2\pi  v \sum_{j=1}^{N-1} \delta(z-z_j) -\frac{(m + \sigma_3)}{(1+ z \bar z)^2}. \]
The index over the whole sphere, including the points associated with the vortices, is
\begin{equation}
 n_+ - n_- = -\frac{1}{2\pi}\int_{S^2} F  = -\bigl[ m +(N-1)v \bigr].\label{eq:index-m}\end{equation}
If the points representing the vortices are excluded the index over the sphere minus $N-1$ points is\footnote{This requires using index theorem for a manifold with boundary (the Atiyah-Patodi-Singer index theorem) where the points are excised. But a careful analysis, using the techniques described in \cite{EGH}, shows that, if $v$ is an integer, the boundary terms arising from small circles surrounding the excised points give no contribution to the index.}
\begin{equation}
 n_+ - n_- = -\frac{1}{2\pi}\int_{S^2 - (N-1)\ \hbox{\small points}} F = - m,\label{eq:index-mtilde}\end{equation}

Zero modes of (\ref{eq:Dzv}) are 
\begin{equation}\chi_-=  \frac{ \bar z^{\bar p} }{(1+z \bar z)^{\frac {m - 1} 2}}
  \prod_{j=1}^{N-1}\frac{(z - z_j)^{-\frac v 2}}{(\bar z - \bar z_j)^{\bar l}},
\label{eq:chilbar}\end{equation}
with $-\frac{v}{2}\ge \bar l$ for regularity at $z_j$.\footnote{When $\bar l \le 0$ this is immediate,
when $\bar l>0$ we invoke 
\[ D_ z \chi_- = 
- 2 \pi \bar l \left[\sum_{j=1}^{N-1}(\bar z - \bar z_j)\delta(\bar z - \bar z_j) \right]\chi_- =0.\]}
Similarly zero modes of (\ref{eq:Dbarzv})
are
\begin{equation}\chi_+= z^p (1+z \bar z)^{\frac {m + 1} 2}
  \prod_{j=1}^{N-1}\frac{(\bar z - \bar z_j)^{\frac v 2}}{(z - z_j)^l}.\label{eq:chil}
\end{equation}

However (\ref{eq:chilbar}) are not all linearly independent,
one could take linear combinations with $\bar z_i$ dependent co-efficients to construct a numerator that has
powers of $\bar z-\bar z_i$ which change $\bar l$, $\bar l$ and $\bar p$ are not independent in
(\ref{eq:chil}). Similarly $l$ and $p$
are not independent in (\ref{eq:chilbar}).  We seek a criterion for constraining
$l$ and $\bar l$ and we shall explore this by looking at the transformation properties under rotations. Of course $SU(2)$ is no longer a symmetry when there are vortices present, but we can still ask how the wave functions (\ref{eq:chilbar}) and (\ref{eq:chil}) change under rotations.

In the presence of vortices the spinor Lie derivatives introduced before (\ref{eq:J-def}) are modified to
\begin{align}
 {\bm J}_+   & = z^2 \partial_z + \partial_{\bar z} + 
\frac v 2 \sum_{j=1}^{N-1} \left( \frac{z^2}{z - z_j}  +  \frac{1}{\bar z - \bar z_j}\right)
        +\frac{(m+ \sigma_3)z}{2}
        \nonumber \\
 {\bm J}_-   & = -{\bar z}^2 \partial_{\bar z} - \partial_z + \frac v 2 \sum_{j=1}^{N-1}\left( \frac{\bar z^2}{\bar z - \bar z_j}  +  \frac{1}{z - z_j}\right)
 + \frac{(m + \sigma_3) \bar z}{2} \label{eq:J-def-vortices} \\  
{\bm J}_3  &= z \partial_z - \bar z \partial_{\bar z}+
 \frac v 2 \sum_{j=1}^{N-1}\left( \frac{z}{z - z_j}  +  \frac{\bar z}{\bar z - \bar z_j}\right)       
+ \frac{(m +\sigma_3)}{2}
        \nonumber
\end{align} 
on the northern hemisphere.
These generate $SU(2)$ at any point on the sphere away from the vortices, but not at the vortices themselves ---
at the vortices there will be delta function singularities that prevent the algebra from closing. 
The algebra is well defined and closes on the
sphere with $N-1$ points removed.

For ${\bm J}_3$
\begin{equation}
  [{\bm J}_3,D_z] = -D_z - 2 \pi v \bar z \sum_{j=1}^{N-1} \delta(z-z_j), 
\qquad [{\bm J}_3,D_{\bar z}]  = D_{\bar z} - 2 \pi v z \sum_{j=1}^{N-1}\delta(z-z_j).\end{equation}
This implies that ${\bm J}_3$ commutes with the Dirac operator on the sphere with $N-1$ points removed.
A short calculation gives
\begin{equation}
{\bm J}_3 \chi_+ =\left(p +\left(\frac v 2 - l\right) \sum_{j=1}^{N-1} \frac{z}{z-z_j} + 2 \pi l \bar z \sum_{j=1}^{N-1}(z-z_j)\delta (z-z_j)  +\frac{m +1}{2}\right)\chi_+,\label{eq:L_3chi_+}
\end{equation}
if $v$ and $l$ are both positive.
Choosing $l=\frac{v}{2}$ results in 
\[
\chi_+ = z^p (1+ z \bar z)^{\frac{m+1}{2}}
\prod_{j=1}^{N-1}\left( \frac{\bar z - \bar z_j}{z - z_j}\right)^{\frac v 2}\]
with
\[
{\bm J}_3 \chi_+ = \left(p + \pi v \bar z \sum_{j=1}^{N-1}\delta(z-z_j) + \frac{m + 1}{2}\right) \chi_+
\]
and $\chi_+$ is an eigenfunction of ${\bf J}_3$ on the punctured sphere with the $N-1$ points removed.

We restrict $p$ to be a non-negative integer, so as to render $\chi_+$ well behaved\footnote{By well behaved we mean that it is finite and, apart from the overall factor of $1/(1+ z \bar z)^{\frac{(|m|-1)}{2}}$, it is a product of a function analytic in $z$
and a function analytic in $\bar z$.} at $z=0$,
and take $m <0$ with $p \le |m|-1$ so that $\chi_+$ is well behaved as $|z|\rightarrow \infty$. 
Then
\begin{equation} \chi_+ = \frac{z^p}{(1+ z \bar z)^{\frac{|m|-1}{2}}}
\prod_{j=1}^{N-1}\left( \frac{\bar z - \bar z_j}{z - z_j}\right)^{\frac v 2}.\label{eq:chi+singular}\end{equation}
Singularities at the points $z_j$ are evident here as the phase of (\ref{eq:chi+singular}) is undefined there when $v\ne 0$.
So the points with the vortices have to be excised from the sphere and the index on the punctured sphere is given by (\ref{eq:index-mtilde}).  There are no normalisable negative chirality zero modes when $m$ is negative, as can be checked explicitly,
so $n_+ = |m|$.
For $\chi_-$ the analysis is similar, except $m >0$ and (\ref{eq:chi+singular}) is complex conjugated.  

Excising points and using (\ref{eq:chi+singular}) for the zero modes may seem natural
from a mathematical point of view but physically it is not satisfactory.
In the flux attachment picture each of the vortices is attached to a particle
and we wish to include all the particles in the dynamics, we do not want to 
remove these points.  We can avoid excising points yet still satisfy the index theorem by choosing $l=-\frac{v}{2}$.  Now
\begin{equation} \chi_+ = z^p ( 1 + z \bar z)^{\frac{m+1}{2}}
\prod_{j=1}^{N-1}(\bar z - \bar z_j)^{\frac v 2}(z - z_j)^{\frac v 2}\label{eq:chi+m}\end{equation}
is well behaved for
$p\ge 0$, $v=2 k \ge  0$, where $k$ is an integer,  and
\begin{equation} p + m  + 1 + 2(N-1) k \le 0 \qquad \Rightarrow \qquad
0 \le p \le -(m  + 1 + 2(N-1) k)=-m'-1,\end{equation}
where $m' = m + 2 k (N-2)$.
Thus $m' \le -1$ for positive chirality zero modes (there are no normalisable negative chirality zero modes for negative $m'$). 
The index is now (\ref{eq:index-m}) and $n_+ = |m'|$. With $0\le p \le |m'|-1$ equation (\ref{eq:chi+m})
then is a complete set for the zero modes,
though they are not eigenstates of ${\bm J}_3$.
$m'$ is the effective magnetic charge the
composite fermions see, since $m'$ and $m$ are both negative $|m'|<|m|$ and the composite fermions move in a weaker field than that generated by the monopole $m$, a consequence of the vortices is that the composite Fermions effectively move in a weakened background field.

The net result is that, if we do not wish to excise the vortices from the sphere, then the number of zero modes for negative $m'$ is $|m'|$ and
\begin{equation}
 \chi_+ = \frac{z^p}{( 1 + z \bar z)^{\frac{|m|-1}{2}}} \prod_{j=1}^{N-1}|z - z_j|^{2 k},\label{eq:chi+m-k}\end{equation}
with $v=2 k>0$ and $m = m' - 2(N-1) k <0$.  The vortices necessarily have even charge and act to oppose 
the background monopole field, thus reducing the effective magnetic field that
the composite fermions see.\footnote{Again a similar analysis for $\chi_-$ changes the sign of $m$, and $m'$ with $v=2 k$ and complex conjugates (\ref{eq:chi+m-k}).}  

A general zero mode is a linear combination,
\begin{equation}
 \chi_+ = \frac{1}{( 1 + z \bar z)^{\frac{|m|-1}{2}}} 
\prod_{j=1}^{N-1}|z - z_j|^{2 k}\sum_{p=0}^{|m'|-1}  a_p z^p ,
\label{eq:chi_+}
\end{equation} 
In the flux attachment picture each of the vortices is attached to a particle.
With $N$ particles the antisymmetrised many-particle wave function is
\[ \chi_+(z_1,\ldots,z_N) = \left(\prod_{i=1}^N \frac{1}{(1+ z_i \bar z_i)^{\frac{|m| -1}{2}}}\right)
\left(\prod_{i<j}^N |z_i - z_j|^{2 k}\right)
\sum_{p_1,\ldots,p_N=0}^{|m'|-1}
a_{p_1 \ldots p_N} z_1^{p_1} \cdots z_N^{p_N},\]
where $a_{p_1 \ldots p_N}$ is anti-symmetric.  The ground state is unique if and only
if $a_{p_1 \ldots a_N}$ is unique (up to an overall constant) and this requires
$|m'|=N$ with $a_{p_1 \ldots a_N} \propto  \epsilon_{p_1 \ldots a_N}$. The unique (un-normalised) ground state is
\begin{equation} 
\chi_+(z_1,\ldots,z_N) = \left(\prod_{i=1}^{N} \frac{1}{(1+ z_i \bar z_i)^{\frac{|m| -1}{2}}}\right)
\prod_{i<j}^{N} |z_i - z_j|^{2 k}(z_i - z_j).\label{eq:fractional-ground-state}
\end{equation}
This is the ground state for a system of non-interacting composite fermions each consisting of an electron bound to a vortex 
of strength $2k$ and 
subject to a background field consisting of a magnetic monopole of charge $m$.  
Wave functions of this form on a plane, and hence with a different geometrical factor, were considered by \cite{G-J} and studied numerically in \cite{FFMS}.

There is an energy gap as before  
and the filling factor is
\[ \nu = \frac{N}{|m|} = \frac{N}{N + 2 k(N-1)}\quad \rightarrow \quad \frac{1}{2 k+1} \qquad \hbox{as} \quad N\rightarrow \infty.\]
The system therefore describes the Laughlin series of the fractional quantum Hall effect.

The vortices can be removed by a singular gauge transformation,
\[\chi_+ \rightarrow e^{-i\Phi}\chi_+, \qquad A \rightarrow A + i e^{i \Phi} d e^{-i\Phi}\]
where the phase $\Phi$ is
\[ \Phi =
  \frac {v} {2 i}  \sum_{i<j}^N\ln
  \left(\frac{\bar z_i - \bar z_j}{z_i - z_j}\right).\]
With $v = 2 k$, the ground state $\chi_+$ (\ref{eq:fractional-ground-state}) gauge transforms to
\begin{equation} \chi_+ \ \rightarrow \ \widetilde \chi_+= \left(\prod_{i=1}^{N} \frac{1}{(1+ z_i \bar z_i)^{\frac{|m| -1}{2}}}\right)
\prod_{i<j}^{N} (z_i - z_j)^{2 k+1}.\label{eq:Laughlin}\end{equation}
This is Haldane's ground state for the quantum effect on a sphere, \cite{Haldane}, 
apart from the geometrical factor $\prod_{i=1}^{N} (1+ z_i \bar z_i)^{-\left(\frac{|m| -1}{2}\right)}$ it
is the Laughlin ground state \cite{Laughlin}.
This is a zero mode for $N$ electrons in a background gauge field for a monopole of charge $m <0$ with
potential
\[  \widetilde A^{(+)} = \frac{i m}{2}\left(\frac{z d \bar z - \bar z d z}{1+z \bar z}\right),
\qquad \widetilde A^{(-)}=\frac{i m}{2} \frac{1}{(1+ z \bar z)}  \left( \frac{d z}{z} - \frac{d \bar z}{\bar z} \right),\]
but it is not unique.  
The most general zero mode for this configuration is
\[ \widetilde \chi_+(z_1,\ldots,z_N) = \left(\prod_{i=1}^N \frac{1}{(1+ z_i \bar z_i)^{\frac{|m| -1}{2}}}\right)
\sum_{p_1,\ldots,p_N=0}^{|m|-1}
a_{[p_1 \ldots p_N]} z_1^{p_1} \cdots z_N^{p_N},\]
The degeneracy is determined by the number of components of the anti-symmetric co-efficients 
$a_{[p_1 \ldots p_N]}$, but now regularity at the S pole only requires  $0\le p \le |m|-1$,
so the degeneracy is 
\[\frac{|m|!}{N!(|m| - N)!} = \frac{\bigl((2 k+1\bigr)N - 2k)!}{ N!\bigl(2 k(N-1)\bigr)!},\]
which diverges exponentially as $N\rightarrow\infty$, for any positive $k$. 
The energy gap is lost and one cannot expect the ground state (\ref{eq:Laughlin}) to be stable under
perturbations.  The introduction of the vortices changes (\ref{eq:Laughlin}) to (\ref{eq:fractional-ground-state}) and stabilizes 
the ground state, it is a singular gauge transformation and so can change the physics.

\section { Conclusions \label{sec:Conclusions}}

Haldane's description of the quantum Hall effect on a sphere has been developed in the context of Fermions on a compact space, allowing the Atiyah-Singer index theorem to be utilised in analysing the ground state of the Hamiltonian which necessarily requires zero modes. Electron wave-functions are cross sections of the $U(1)$ bundle associated with the monopole.
For a single electron in the field of a magnetic monopole of charge $m$ (in magnetic units with $\frac {e^2} {h}=1$) the number of zero-modes, and hence the degeneracy of the ground state, is limited by the index theorem to $|m|$. For a system of $N$ particles Fermi statistics then gives the unique ground state (\ref{eq:Haldane}) if and only of $N=|m|$
and the  filling factor is $\nu =1$.
The uniqueness, and hence stability, of the Haldane ground state wave function  for the integer quantum Hall effect (which is the same as the Laughlin ground state function except for a geometric factor) is seen to be a consequence of the index theorem which limits the dimension of the space of zero modes.

The fractional quantum Hall effect can be studied in the composite Fermion scenario by viewing a monopole of charge $m$ to be $|m|$ individual monopoles of charge $\pm 1$ and promoting some of the Dirac strings associated with these monopoles to be statistical gauge field vortices which bind to electrons, forming composite Fermions. The vortices necessarily reduce the magnitude of the background magnetic field seen by the composite fermions and the index theorem again dictates that the degeneracy of the ground state is finite.  Apart from the usual geometrical factor on the sphere the ground state wave-function is a product of a holomorphic and an anti-holomorphic field 
if and only if the vortices are of strength $2 k$ with $k$ an integer.
The ground state wave-function for a system consisting of $N$ composite fermions (\ref{eq:fractional-ground-state}) is then unique if and only if the filling factor is $\frac{1}{2 k+1}$, for large $N$. Removing the vortices by use of a singular gauge transformation then gives the Laughlin ground state for the fractional quantum Hall in the Laughlin series, again apart from a geometrical factor.

It would be interesting to apply similar techniques to the higher dimensional quantum Hall effect \cite{Zhang+Hu}, \cite{Karabali+Nair} in which $S^2$ is replaced by $S^4$ and the spinors are cross-sections of an $SU(2)$ bundle, but we leave that to future work.

\appendix

\section{Vortices on a sphere} 

A vortex of strength $v$ at the N-pole of the sphere, $z=0$, with flux $2\pi v$ out of the sphere, is described by a magnetic potential
\begin{equation}
a=\frac{v}{2 i} \left[ \frac{d z}{z} 
- \frac{d \bar z}{\bar z}\right].\label{eq:vortex-N}
\end{equation}
Then $d a=0$ provided $z\ne 0$ but, if we isolate the N-pole by surrounding it by a small circle $S^1_\epsilon$ of radius $\epsilon$ 
centred on $z=0$, 
\begin{equation}
\int_{S^1_\epsilon} a =
 \frac{v}{2 i}\int_{S^1_\epsilon} \frac{d z}{ z} 
-\frac{v}{2 i}\int_{S^1_\epsilon} \frac{d \bar z}{\bar z}
=   \pi  v  - \pi (- v)= 2\pi v.\label{eq:vortex-S1}
\end{equation}
Thus (\ref{eq:vortex-N}) describes a point vortex of strength $v$ at the N-pole,
$f = da$ is a $\delta$-function at $z=0$,
which can be represented by
\[\partial_{\bar z}\left(\frac{1}{z}\right)=
  \partial_{z}\left(\frac{1}{\bar z}\right) = 2\pi \delta(z).\] 
However this potential also gives an anti-vortex, of strength $-v$, at the S pole:
the anti-podal point is given by $z \rightarrow \frac{1}{\bar z}$, which sends $a\rightarrow -a$. This is perhaps clearer using polar co-ordinates, $(\theta,\phi)$, 
in which 
\[ a = v d \phi\]
which represents an infinite straight flux tube in 3-dimensions, threading through both the N-pole and the S-pole of the sphere.  The total flux through the sphere 
arising from $f = da$ is zero.

The position of the vortex through the N-pole can be moved around by using
\begin{equation}
a=\frac{v}{2 i} \left[ \frac{d z}{(z-z_1)} 
- \frac{d \bar z}{(\bar z - \bar z_1)}\right],\label{eq:vortex-A}
\end{equation}
representing a vortex of strength $v$ through the point $z_1$, but there is still
an anti-vortex through the S-pole for any finite $z_1$.
However the vortex at the S-pole can be removed by adding a uniform magnetic field with a semi-infinite solenoid threading the S pole and terminating at the centre of the sphere,
\begin{equation}
a=\frac{v}{2 i} \left[ \frac{d z}{(z-z_1)} 
- \frac{d \bar z}{(\bar z - \bar z_1)}\right] 
+\frac{v}{2 i}\left(\frac{z d \bar z - \bar z d z}{1+z \bar z}\right),\label{eq:solenoid}
\end{equation}
giving the field strength
\[ f=d a = i\left(2 \pi  v \delta(z_1)  + \frac{v}{2}\frac{1}{(1+ z \bar z)^2}\right) d z \wedge d \bar z \] 
This is perfectly regular at the S pole and represents a magnetic monopole of charge $-v$ at the centre of the sphere together with a point vortex of strength $v$ at $z_1$,
the total flux is zero (figure 1). It is actually like a Dirac monopole with it's accompanying
string threading the sphere at $z_1$, but a Dirac string is a gauge artifact, a vortex is not.

\begin{figure}[h]
\centerline{\includegraphics{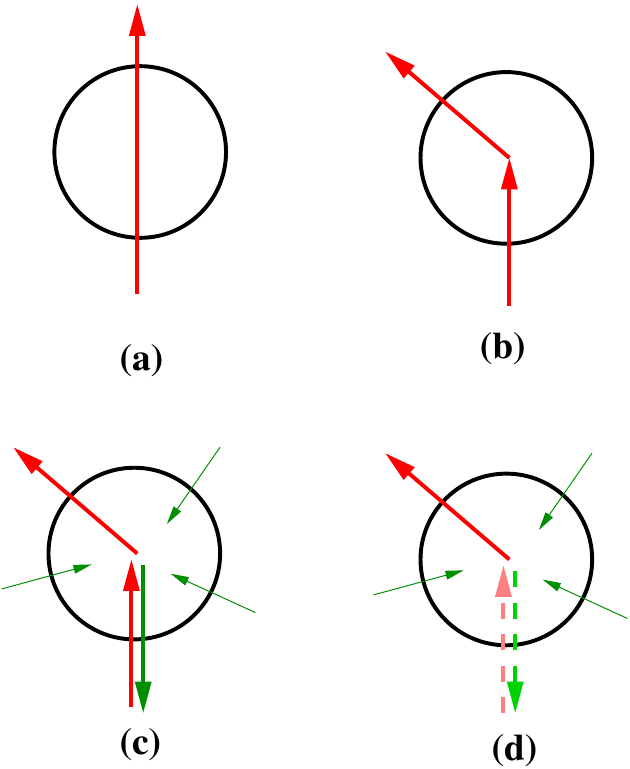}}
\caption{A vortex threading the sphere. (a) shows the simple vortex in (\ref{eq:vortex-N}) piercing the sphere at the north and south poles; 
(b) shows the vortex in (\ref{eq:vortex-A}), piercing the sphere at $z_1$ and the south pole; (c) shows the combination of the vortex in (b) combined with a Dirac monopole of charge $-1$ uniformly distributed on the sphere together with its accompanying string through the south pole; (d) the Dirac string and the vortex through the south pole cancel leaving a uniform monopole field with a vortex at $z_1$.  The total flux through the sphere in (d) is zero --- the Dirac string has been moved  from the south pole to the point $z_1$.}
\end{figure}

If there are $N-1$ vortices all of the same strength $v$ positioned at $z_j$ then the fields are simply added:
\begin{equation}a=\frac{v}{2 i} \sum_{j=1}^{N-1}\left( \frac{d z }{z-z_j}
-  \frac{d \bar z}{\bar z-\bar z_j}\right)
-\frac{i (N-1) v}{2}\left(\frac{z d \bar z - \bar z d z}{1+z \bar z}\right).\label{eq:a}
\end{equation}
The corresponding field strength is
\[ f = d a =i\left( 2 \pi v \sum_{j=1}^{N-1}\delta(z-z_j) -
\frac{(N-1)v}{(1 + z \bar z)^2} \right)  dz \wedge \bar dz\]
and
\[ \int_{S^2} f =0.\]
If in addition a background monopole field with charge $m'$ is present then the total gauge potential on the northern hemisphere is
\begin{equation} A^{(+)}  = \frac{v}{2 i} \sum_{j=1}^{N-1}\left( \frac{d z }{z-z_j}
-  \frac{d \bar z}{\bar z-\bar z_j}\right)
+ \frac{i m}{2}\left(\frac{z d \bar z - \bar z d z}{1+z \bar z}\right),\label{eq:A+}
\end{equation}
where $m=m' - (N-1)v$,
and the field strength is
\[ F = d A^{(+)} = i\left(2 \pi v \sum_{j=1}^{N-1} \delta(z-z_j) + 
\frac{m}{2 (1+z \bar z)^2}\right)dz \wedge d \bar z.\]
On the southern hemisphere we take the potential to be
\begin{align} A^{(-)} 
&= \frac{v}{2 i} \sum_{j=1}^{N-1}\left( \frac{d z }{z-z_j}
-  \frac{d \bar z}{\bar z-\bar z_j}\right)
- \frac{i (N-1)v (z d \bar z - \bar z d z)}{2(1+z \bar z)}
+\frac{i m'}{2} \frac{1}{(1+ z \bar z)}  \left( \frac{d z}{z} - \frac{d \bar z}{\bar z} \right)\nonumber\\
&= \frac{v}{2 i} \sum_{j=1}^{N-1}\left( \frac{d z }{z-z_j}
-  \frac{d \bar z}{\bar z-\bar z_j}\right)
+ \frac{i}{2} \left( \frac{m}{(1+ z \bar z)}  +  (N-1) v \right) 
  \left( \frac{d z}{z} - \frac{d \bar z}{\bar z} \right),\label{eq:A-}
\end{align}
which is perfectly well defined as $|z|\rightarrow \infty$. Again
\[ F = d A^{(-)} = i\left(2 \pi v \sum_{j=1}^{N-1} \delta(z-z_j) + 
\frac{m}{2 (1+z \bar z)^2} \right)dz \wedge d \bar z.\]
The total flux is
\[ \int_{S^2} F= 2\pi \bigl[m + (N-1)v \bigr],\]
see figure 2.

\begin{figure}[h]
\centerline{\includegraphics{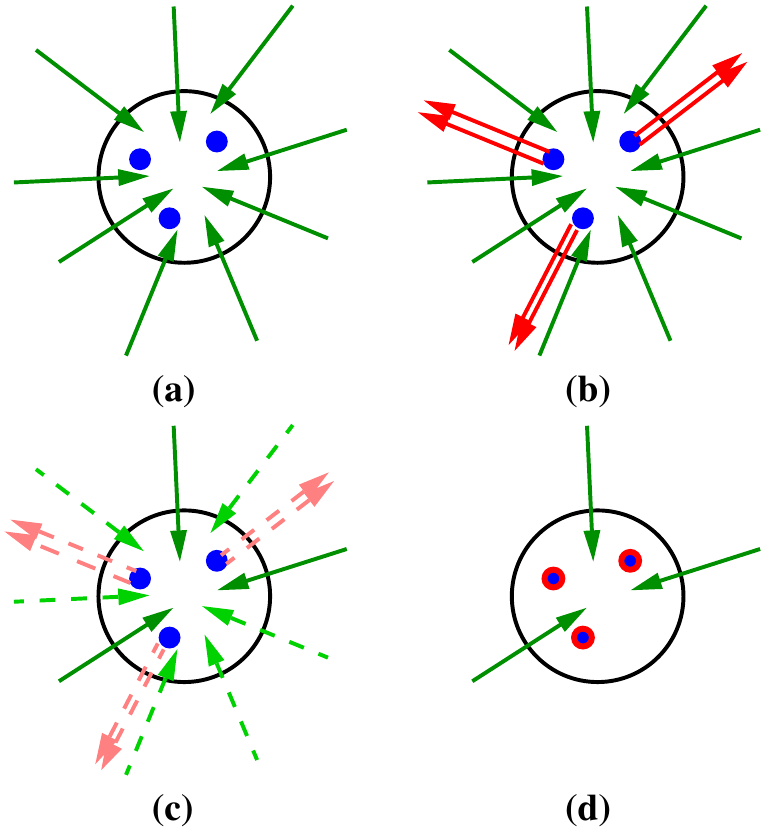}}
\caption{Composite fermions. (a) represents three electrons in a uniform background flux with total magnetic charge $m=-9$, giving filling factor $\frac{1}{3}$;
(b) six Dirac strings are promoted to be real vortices and attached to the electrons in pairs; (c) the total magnetic flux is now $m'=-3$; (d) the resulting configuration consists of three composite fermions in a field of strength $-3$ giving an effective filling factor 1.}
\end{figure}

What we have done here is taken $|m|$ monopoles each of charge $\pm 1$
(depending on the sign of $m$) and promoted the Dirac strings on 
$N-1$ of them to be real vortices at $z_i$, but leaving $|m'|$ of them as Wu-Yang monopoles, for which the Dirac string is a gauge artifact.
The configuration is indistinguishable from that of a monopole of charge $m$
together with $N-1$ vortices.

\end{document}